\documentclass[aps,prb,manuscript,amsmath,superscriptaddress]{revtex4}
\usepackage{graphicx}
\usepackage{color}
\usepackage{epstopdf}
\usepackage{xcolor}

\begin{document}

\title{Individual Cr atom in a semiconductor quantum dot: optical addressability and spin-strain coupling}

\author{A. Lafuente-Sampietro}
\affiliation{Universit\'{e} Grenoble Alpes, Institut N\'{e}el, F-38000 Grenoble, France}
\affiliation{CNRS, Institut N\'{e}el, F-38000 Grenoble, France}
\affiliation{Institute of Material Science, University of Tsukuba, Japan}

\author{H. Utsumi}
\affiliation{Institute of Material Science, University of Tsukuba, Japan}

\author{H. Boukari}
\affiliation{Universit\'{e} Grenoble Alpes, Institut N\'{e}el, F-38000 Grenoble, France}
\affiliation{CNRS, Institut N\'{e}el, F-38000 Grenoble, France}

\author{S. Kuroda}
\affiliation{Institute of Material Science, University of Tsukuba, Japan}

\author{L. Besombes}\email{lucien.besombes@grenoble.cnrs.fr}
\affiliation{Universit\'{e} Grenoble Alpes, Institut N\'{e}el, F-38000 Grenoble, France}
\affiliation{CNRS, Institut N\'{e}el, F-38000 Grenoble, France}

\begin{abstract}
We demonstrate the optical addressability of the spin of an individual Chromium atom (Cr) embedded in a semiconductor quantum dot. The emission of Cr-doped quantum dots and their evolution in magnetic field reveal a large magnetic anisotropy of the Cr spin induced by local strain. This results in the zero field splitting of the 0, $\pm$1 and $\pm$2 Cr spin states and in a thermalization on the magnetic ground states 0 and $\pm$1. The observed strong spin to strain coupling of Cr is of particular interest for the development of hybrid spin-mechanical devices where coherent mechanical driving of an individual spin in an oscillator is needed. The magneto-optical properties of Cr-doped quantum dots are modelled by a spin Hamiltonian including the sensitivity of the Cr spin to the strain and the influence of the quantum dot symmetry on the carrier-Cr spin coupling.
\end{abstract}

\maketitle

In the last years, the development of semiconductor based nano-electronics has driven intensive studies of individual spin dynamics and control in nano-structures. It has been shown that the electrical and the optical properties of a semiconductor quantum dot (QD) can be used to control the spin of individual carriers \cite{Greilich2006,Press2008}, ensemble of nuclei \cite{Urbaszek2013} as well as individual \cite{Besombes2004,LeGall2011,Kudelski2007,Kobak2014} or pairs \cite{Besombes2012,Krebs2013} of magnetic atoms. In contrast to NV centers in diamond or molecular magnets, semiconductor QDs containing individual spins, and in particular a localized spin on a single atom, may be incorporated in conventional semiconductor devices \cite{Pierre2010} to add new quantum functionalities.

Among the variety of magnetic transition elements that can be incorporated in semiconductors, Chromium (Cr) is of particular interest. Cr is incorporated in II-VI semiconductors as Cr$^{2+}$ ions \cite{Vallin1974} (electronic configuration [Ar]3$d^44s^04p^0$) carrying a localized electronic spin S=2 and an orbital momentum L=2. Moreover, 90 $\%$ of Cr isotopes have no nuclear spin limiting the number of spin states to five. The non-zero orbital momentum of Cr and spin-orbit coupling result in a large sensitivity of its spin to local strain. For such magnetic atom, static strain could be used to control the magnetic anisotropy and thus influence its spin memory \cite{Oberg2014,Lin2015}. This large spin to strain coupling also makes Cr a very promising spin $qubit$ for the realization of hybrid spin-mechanical systems in which the motion of a microscopic mechanical oscillator would be coherently coupled to the spin state of a single atom \cite{Tessier2014,Ovar2014}.

We demonstrate here for the first time that the spin of an individual Cr atom can be detected optically. We show that the spin states of a Cr embedded in a CdTe/ZnTe QD, are split by a large magnetic anisotropy induced by biaxial strain in the plane of the QDs. At low temperature (T=5K), only the low energy states corresponding to the Cr spin S$_z$=0 and S$_z$=$\pm$1 are detected by photoluminescence (PL). In some of the QDs, a significant mixing between dark and bright excitons attributed to the reduced symmetry of the dots is observed. The magnetic field dependence of the PL of an exciton in the exchange field of a Cr spin is analyzed. Magneto-optical properties of Cr-doped QDs can be modelled by a spin Hamiltonian including the strain induced fine structure of the magnetic atom, the exchange coupling with the carriers and the influence of the reduced symmetry of the QDs on the electron-hole exchange interaction and on the valence band.

Cr atoms are randomly introduced in CdTe/ZnTe self-assembled QDs grown by Molecular Beam Epitaxy on ZnTe (001) substrates following the procedure described in ref.\cite{Wojnar2011}. The amount of Cr is adjusted to optimize the probability to detect QDs containing 1 or a few Cr atoms. The emission of individual QDs, induced by optical excitation with a dye laser tuned on resonance with an excited state of the dots \cite{Besombes2014}, is studied in magnetic fields (up to 11 T) by optical micro-spectroscopy in Faraday configuration.

\begin{figure}[hbt]
\includegraphics[width=3.0in]{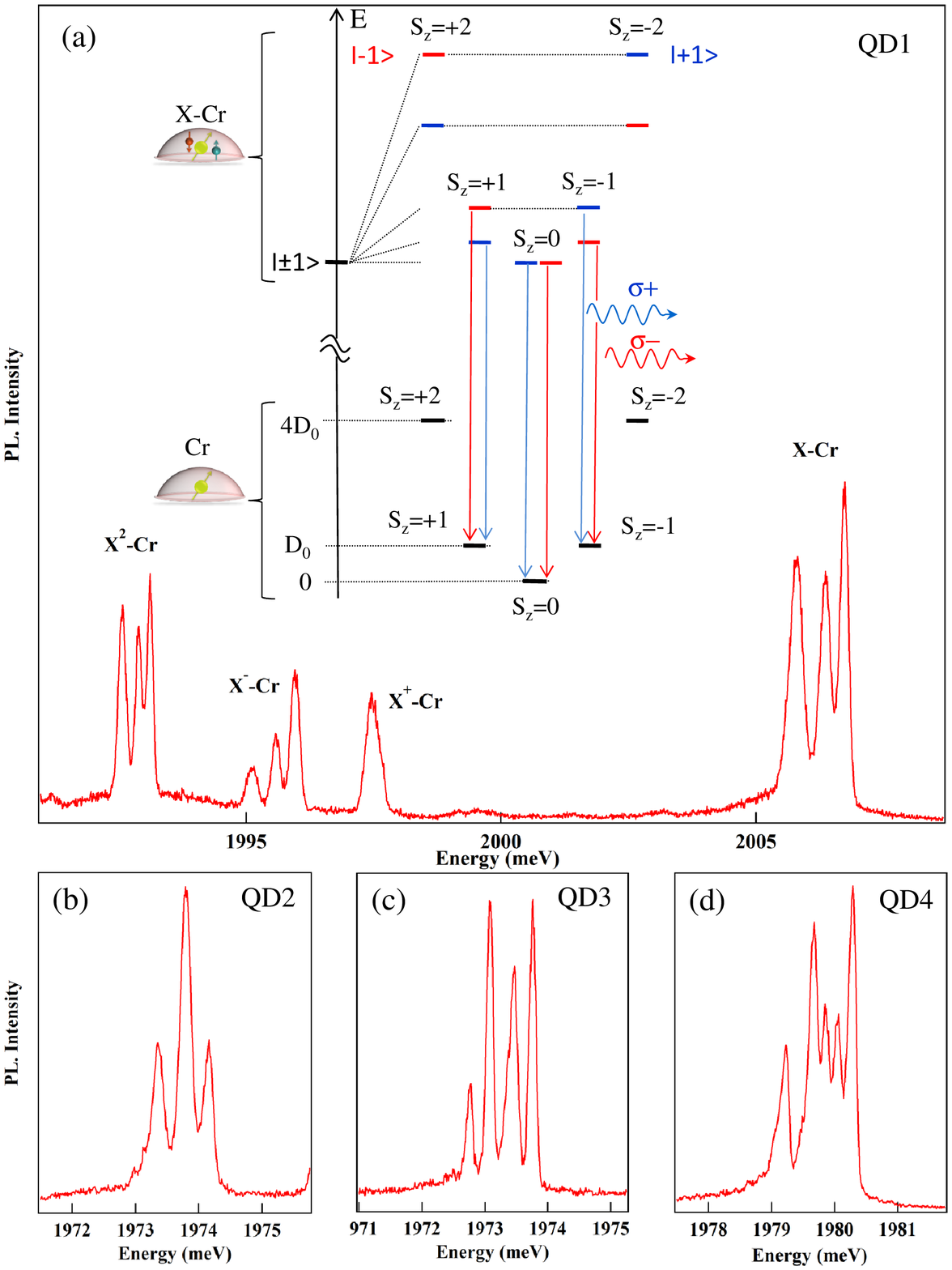}
\caption{(a) Low temperature (T=5K) PL of a QD (QD1) doped with a single Cr atom showing the emission from the neutral exciton (X-Cr), positively and negatively charged excitons (X$^+$-Cr and X$^-$-Cr) and biexciton (X$^2$-Cr). The energy levels of the ground state (Cr), the bright exciton states ($|\pm1\rangle$) coupled to the spin of a Cr (X-Cr) and dominant PL transitions ($\sigma$+, $\sigma$-) are illustrated in the inset. (b-d) PL of X-Cr in three other QDs (QD2, QD3 and QD4).}
\label{Fig1}
\end{figure}

The low temperature (T=5K) PL of the neutral exciton (X-Cr), positively charged (X$^+$-Cr) and negatively charged (X$^-$-Cr) excitons and biexciton (X$^2$-Cr) of an individual Cr-doped QD (QD1) are reported in Fig.~\ref{Fig1}(a). Three emission lines are observed for the neutral species (X and X$^2$) and the positively charged excitons. The relative intensities of the lines and their splitting change from dot to dot as illustrated in Fig.~\ref{Fig1} (b-d). For some of the dots, a splitting of the central line is observed for X-Cr and X$^2$-Cr and an additional line appears on the low energy side of the X-Cr spectra (see QD3 and QD4 Fig.~\ref{Fig1} (c-d)). All these features result from the exchange coupling of the electron and hole spins with a single Cr spin.

In a II-VI semiconductor, the orbital momentum of the Cr connects the spin of the atom to its local strain environment through the modification of the crystal field and the spin-orbit coupling. For biaxial strain in the (001) plane, the ground state of a Cr spin is split by a strain induced magnetic anisotropy term ${\cal H}_{Cr,\varepsilon_\parallel}=D_0S^2_z$ \cite{Sup}. It was deduced from electron paramagnetic resonance of bulk Cr-doped CdTe that $D_0$ is positive for compressive biaxial strain \cite{Vallin1974}. In a self-assembled CdTe/ZnTe QD with large in-plane strain, the Cr spin energy levels are split with S$_z$=0 at low energy (Scheme in Fig.~\ref{Fig1}(a)). A value of $D_0$ in the 1 meV range can be expected for a CdTe layer strained on a ZnTe substrate \cite{Vallin1974,Sup}.

When an electron-hole (e-h) pair is injected in a Cr-doped QD, the bright excitons are split by the exchange interaction between the spins of Cr and carriers. In flat self-assembled QDs, the heavy-holes and light-holes are separated in energy by the biaxial strain and the confinement. In a first approximation, the ground state in such QD is a pure heavy-hole (J$_z$=$\pm$3/2) exciton and the exchange interaction with the Cr spin S is described by the spin Hamiltonian ${\cal H}_{c-Cr}=I_{eCr}\vec{S}\cdot\vec{\sigma}+I_{hCr}S_zJ_z$, with $\vec{\sigma}$ the electron spin and J$_z$ the hole spin operator. I$_{eCr}$ and I$_{hCr}$ are, respectively, the exchange integrals of the electron and the hole spins with the Cr spin. These exchange energies depend on the exchange constant of the $3d$ electrons of the Cr with the carriers in CdTe and on the overlap of the Cr atom with the confined carriers. The exchange interaction of the Cr spin is ferromagnetic for both electron and hole spins in common II-VI semiconductors and a typical exchange constant 4 to 5 times larger for the holes than for the electrons is also expected in CdTe \cite{Mac1996,Herbich1998}.

\begin{figure}[hbt]
\includegraphics[width=3.0in]{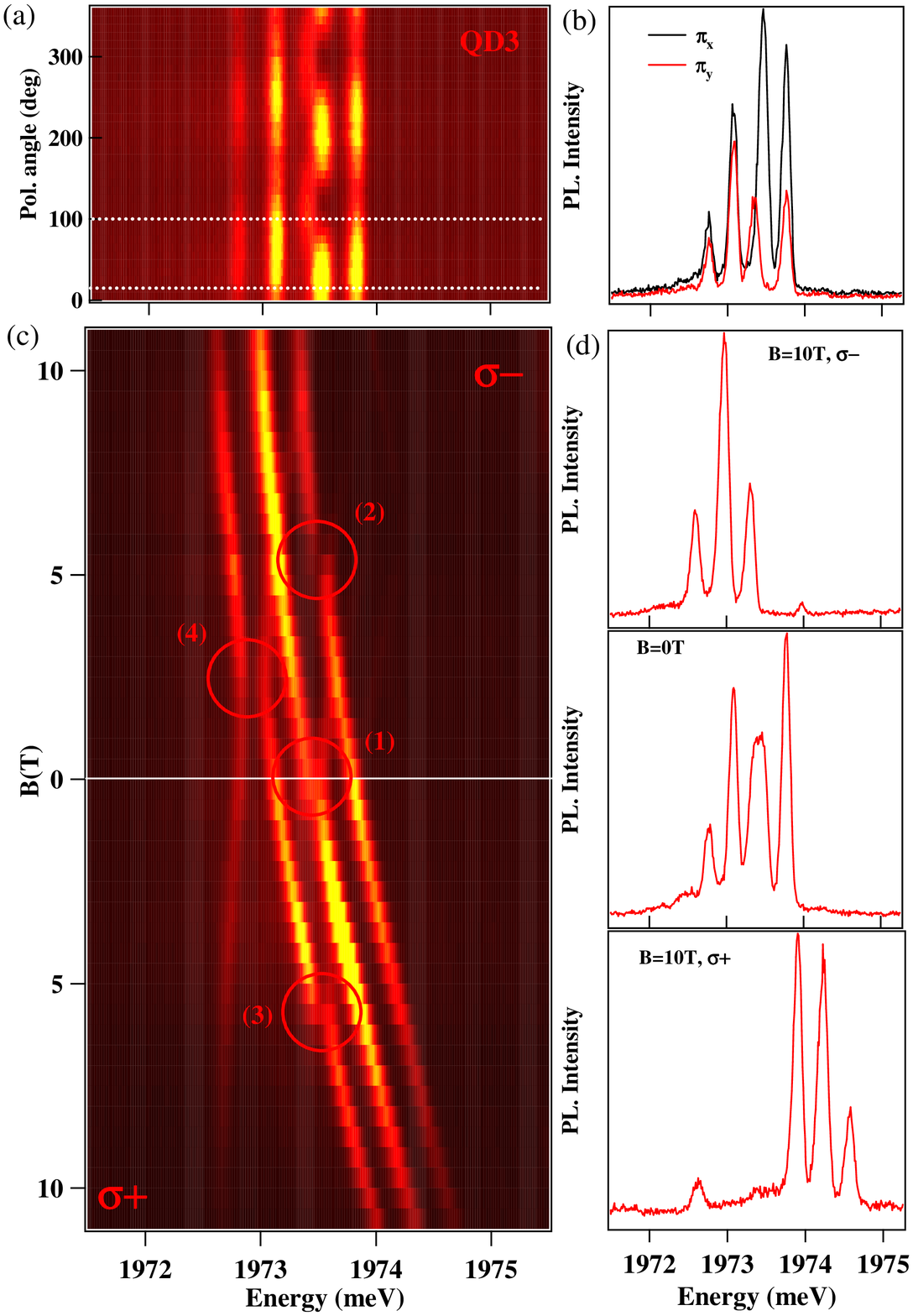}
\caption{(a) Linear polarization PL intensity map of X-Cr in QD3. (b) Corresponding linearly polarized PL spectra. (c) PL intensity map of X-Cr in QD3 versus magnetic field (B$_z$). (d) Corresponding circularly polarized PL spectra recorded at B$_z$=0T and B$_z$=10T.}
\label{Fig2}
\end{figure}

For highly strained CdTe/ZnTe QDs with a weak hole confinement, the strain induced energy splitting of the Cr spin $D_0S^2_z$ is much larger than the exchange energy with the confined carriers ($D_0\gg |I_{hCr}|>|I_{eCr}|$). The exchange interaction with the exciton acts as an effective magnetic field which further splits the Cr spins states S$_z$=$\pm$1 and S$_z$=$\pm$2. The resulting X-Cr energy levels are presented in Fig.~\ref{Fig1}(a). The exciton recombination does not affect the Cr atom and its spin is conserved during the optical transitions. Consequently, the large strained induced splitting of the Cr spin is not directly observed in the optical spectra. However, at low temperature, the Cr spin thermalize on the low energy states S$_z$=0 and S$_z$=$\pm$1. This leads to a PL dominated by three contributions: A central line corresponding to S$_z$=0 and the two outer lines associated with S$_z$=$\pm$1 split by the exchange interaction with the carriers.

\begin{figure}[hbt]
\includegraphics[width=3.25in]{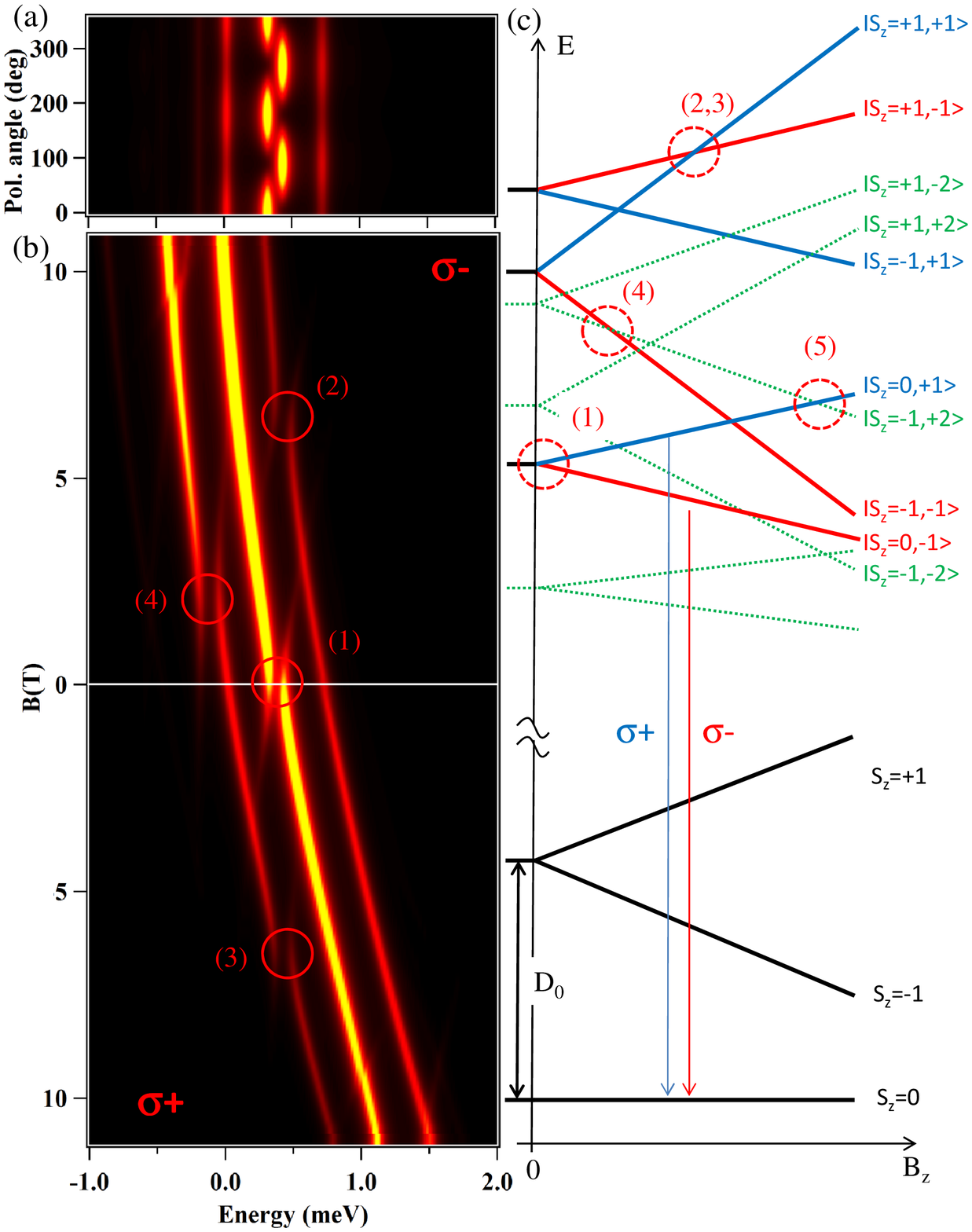}
\caption{(a) Calculated linear polarization PL intensity map of X-Cr at zero field and (b) calculated circularly polarized magnetic field dependence. Details of the model and parameters are listed in table II of the supplemental material. (c) Scheme of the magnetic field dependence of the energy levels of the low energy Cr spin states S$_z$=0 and S$_z$=$\pm$1 and corresponding bright ($|+1\rangle$ blue, $|-1\rangle$ red) and dark ($|\pm2\rangle$ green) X-Cr energy levels.}
\label{Fig3}
\end{figure}

The structure of the energy levels in Cr-doped QDs is confirmed by the evolution of the PL spectra in magnetic field. The circularly polarized PL of QD3 under a magnetic field applied along the QD growth axis is presented in Fig.~\ref{Fig2} together with an analysis of the linear polarization at zero field. The central line (S$_z$=0) is split and linearly polarized along two orthogonal directions (Fig.~\ref{Fig2}(a) and (b)). As in non-magnetic QD, this results from a coupling of the two bright excitons $|\pm1\rangle$ by (i) the short range e-h exchange interaction in the presence of valence band mixing and/or (ii) the long-range e-h exchange interaction in a QD with an in-plane shape anisotropy. This anisotropic e-h exchange energy mixes the bright exciton associated with the same Cr spin state, inducing an extra splitting between them. The mixing is maximum for the central pair of bright exciton (S$_z$=0) which are initially degenerated. The outer lines are also slightly linearly polarized but the influence of the e-h exchange interaction is attenuated by the initial splitting of the $|\pm1\rangle$ excitons induced by the exchange interaction with the Cr spin S$_z$=$\pm1$. As illustrated in Fig.~\ref{Fig3}(c), the Zeeman energy of the exciton under magnetic field can compensate the exciton splitting induced by the exchange interaction with the Cr \cite{Leger2005}. For QD3, this results in an anti-crossing of $|+1\rangle$ and $|-1\rangle$ excitons due to the e-h exchange interaction around B$_z$=6 T observed both in $\sigma$+ and $\sigma$- polarizations (anti-crossing (2) and (3) in Fig.~\ref{Fig2}(c)).

In some of the Cr-doped QDs, an additional line appears on the low energy side of the PL spectra at zero magnetic field (line at 1972.7 meV for QD3 and at 1979.1 meV for QD4 in Fig.~\ref{Fig1}). An anti-crossing of this line with the bright excitons is observed under B$_z$ in $\sigma$- polarization (anti-crossing (4) in Fig.~\ref{Fig2}). As illustrated in Fig.~\ref{Fig3}(c) this anti-crossing arises from a mixing of the bright and dark excitons interacting with the same Cr spin state. Observed in $\sigma$- polarization, it corresponds to the mixing of the exciton states $|-1\rangle$ and $|+2\rangle$ coupled to the Cr spin S$_z$=-1. This dark/bright exciton coupling $\delta_{12}$ is induced by the e-h exchange interaction in a confining potential of reduced symmetry (lower than C$_{2v}$) \cite{Zielinski2015}. In such symmetry, the dark excitons acquire an in-plane dipole moment which lead to possible optical recombination at zero magnetic field \cite{Bayer2002} as observed in QD3 and QD4. The oscillator strength of this "dark exciton" increases as the initial splitting between $|-1\rangle$ and $|+2\rangle$ excitons is reduced by the magnetic field (Fig.\ref{Fig3}(c)).

We calculated the magneto-optic behavior of Cr-doped QDs like QD3 by diagonalizing the complete Hamiltonian of the e-h-Cr system \cite{Sup}. We considered the general case of QDs with a symmetry lower than C$_{2v}$ (truncated ellipsoidal lens for instance \cite{Zielinski2015}), and took into account the influence of this reduced symmetry on the valence band and on the e-h exchange interaction. The population of the X-Cr spin states split by the large magnetic anisotropy and the carriers-Cr exchange interaction is described by a spin effective temperature T$_{eff}$. The results of the model obtained with T$_{eff}$=25K, D$_0$=2.5 meV and an electron-Cr (hole-Cr) exchange interaction I$_{eCr}$=-70$\mu$eV (I$_{hCr}$=-280 $\mu$eV) are reported in Fig.~\ref{Fig3} (parameters not specific to Cr-doped QDs are listed in table II of the Supplemental Material). The PL of X-Cr at zero field and its evolution in magnetic field can be qualitatively reproduced. In particular, the description of the spin states occupation by T$_{eff}$ is sufficient to reproduce the observed emission from the three low energy X-Cr levels (Cr spin states S$_z$=0 and S$_z$=$\pm$1). The splitting of the central line at zero field (anti-crossing (1)) and the anti-crossings under magnetic field (anti-crossings (2) and (3) around B$_z$=6T for the Cr spin states S$_z$=+1 and anti-crossings (4) with the dark exciton around B$_z$=2T) are also well reproduced by the model.

The magnetic anisotropy D$_0$ cannot be precisely extracted from the PL spectra. However, a too large value would produce a smaller PL intensity of the sates S$_z$=$\pm$1 than observed experimentally. In addition, for D$_0$ $<$ 2.25 meV, an anti-crossing due to an electron-Cr flip-flop controlled by I$_{eCr}$, labelled (5) in Fig.\ref{Fig3}(c), would appear below B$_z$=11T on the central line in $\sigma$+ polarization.

\begin{figure}[hbt]
\includegraphics[width=3.25in]{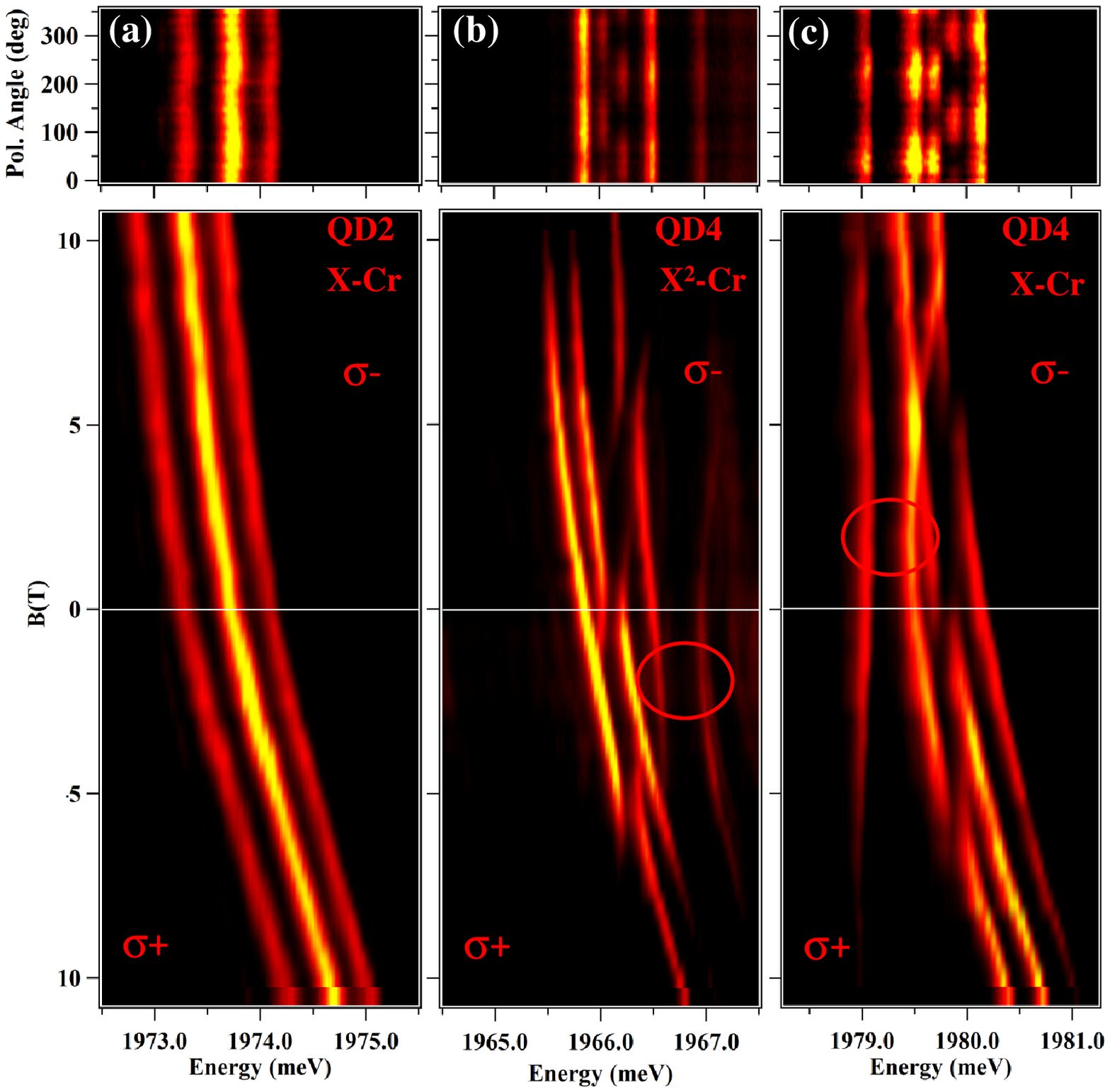}
\caption{Linear polarization intensity map (top panel) and intensity map of the longitudinal magnetic field dependence of the emission (bottom panel) of (a) X-Cr in QD2, (b) X$^2$-Cr in QD4 and (c) X-Cr in QD4.}
\label{Fig4}
\end{figure}

To illustrate the influence of the QD symmetry on the magneto-optical properties of X-Cr, we compare in Fig.~\ref{Fig4} the emission of a QD with cylindrical symmetry (QD2, Fig.~\ref{Fig4}(a)) and a QD with low symmetry (QD4, Fig.~\ref{Fig4}(c)). For QD2, only three unpolarized lines are observed at zero field. Under magnetic field, each line is split and their energy follow the Zeeman and diamagnetic shift of the exciton. In QDs with a reduced symmetry such as QD4, a large contribution from the dark exciton is observed in the PL at zero field and the emission presents linearly polarized components. The characteristic bright excitons mixing induced by the e-h exchange interaction are observed in both circular polarization under magnetic field.

Investigating both the biexciton and the exciton in the same Cr-doped QD, we can also analyze the impact of the carrier-Cr interaction on the fine structure of the Cr spin. The magnetic field dependence of X$^2$-Cr and X-Cr emissions in QD4 are presented as a contour plot in Fig.~\ref{Fig4}(b) and (c) respectively. The PL under magnetic field of X-Cr and X$^2$-Cr present a mirror symmetry. In particular, the dark/bright exciton mixing observed around B$_z$=2.5T on the low energy side of the PL in $\sigma-$ polarization for X-Cr is observed on the high energy side in $\sigma+$ polarization for X$^2$-Cr (circles in Fig.~\ref{Fig4}(b) and (c)).

If one consider the ground state of X$^2$ as a spin-singlet (total spin 0), it cannot be split by the magnetic field or the spin interaction part of the carriers-Cr Hamiltonian. The creation of two excitons in the QD cancels the exchange interaction with the Cr atom. Thus, the PL of  X$^2$-Cr is controlled by the final state of the optical transitions, i.e. the eigenstates of X-Cr, resulting in the observed mirror symmetry in the PL spectra. However, in some of the QDs, the X$^2$-Cr emission slightly deviates from this simple picture: a smaller energy splitting is observed for X$^2$-Cr compared to X-Cr (see X-Cr and X$^2$-Cr in QD1 and Supplemental Materials). This shows that there is an interaction of X$^2$ with the Cr atom. It could result from a perturbation of the carriers' wave function by the interaction with the magnetic atom \cite{Besombes2005,Trojnar2013} or a modification the local electric field which controls the Cr fine structure. Further investigations are required to explain this effect that goes beyond the purpose of this article.

To conclude, we demonstrated that the spin of a Cr atom in a semiconductor can be probed optically. The fine structure of the Cr is dominated by a magnetic anisotropy induced by strain in the plane of the QDs. The large spin to strain coupling of Cr, two orders of magnitude larger than for magnetic elements without orbital momentum (NV centers in diamond \cite{Barfuss2015}, Mn atoms in II-VI semiconductors \cite{Lafuente2015}) suggests some possible development of coherent mechanical spin-driving of an individual magnetic atom in a nano-mechanical oscillator. This new single spin system should allow, at low temperature, to enter some coupling regimes dominated by quantum coherent dynamics not reached until now in hybrid spin-mechanical devices.

\begin{acknowledgements}

The authors acknowledge financial support from the Labex LANEF for the Grenoble-Tsukuba collaboration. This work was realized in the framework of the Commissariat \`{a}  l'Energie Atomique et aux Energies Alternatives (Institut Nanosciences et Cryog\'{e}nie) / Centre National de la Recherche Scientifique (Institut N\'{e}el) joint research team NanoPhysique et Semi-Conducteurs.

\end{acknowledgements}

\newpage

\begin{center}

{\large{\bf Supplemental Material to "Individual Cr atom in a semiconductor quantum dot: optical addressability and spin-strain coupling"}}

\end{center}

In this supplemental material, we first describe the influence of strain on the fine structure of a Cr atom embedded in a II-VI zinc-blende semiconductor. We then present the complete electron-hole-Cr spin effective Hamiltonian used to model a low symmetry singly Cr-doped II-VI quantum dot. We finally present additional data on Cr-doped quantum dots: the temperature dependence of the PL of X-Cr and some statistics on the energy splitting of X-Cr and X$^2$-Cr.

\section{Cr energy levels in a II-VI semiconductor}

\begin{figure}[hbt]
\includegraphics[width=3.0in]{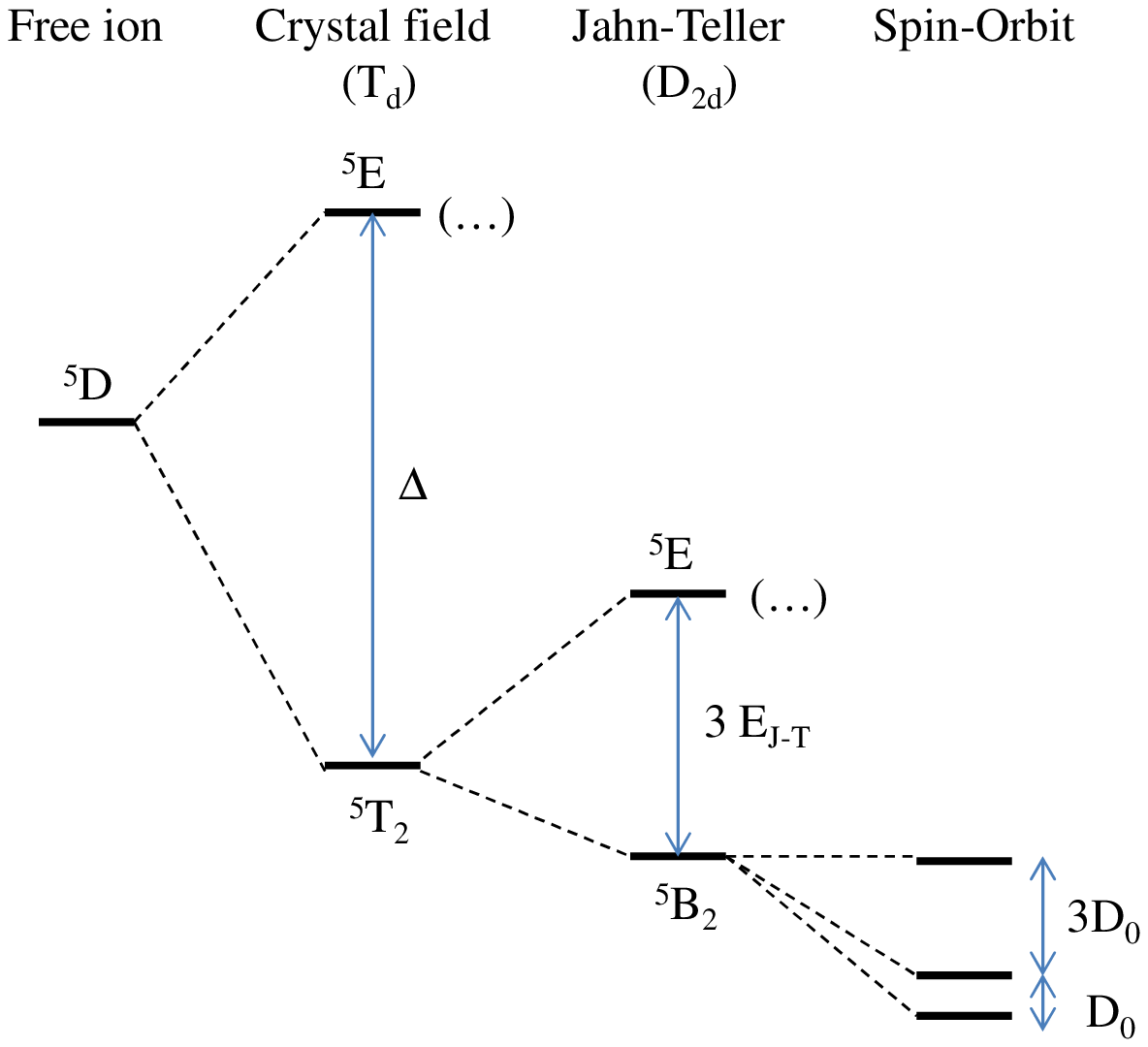}
\caption{Scheme of the energy level splitting of Cr$^{2+}$ at a cation site in II-VI compounds having zinc blende structure (T$_d$) with a crystal field parameter $\Delta$, a Jahn-Teller energy E$_{J-T}$ and a spin-orbit level spacing D$_0$.}
\label{FigS1}
\end{figure}

Cr atoms are incorporated into II-VI semiconductors as Cr$^{2+}$ ions on cation sites forming a deep impurity level. The ground state of a free Cr$^{2+}$ is $^{5}$D with the orbital quantum number L=2 and a spin S=2 yielding a 25-fold degeneracy. In the crystal field of T$_{d}$ symmetry of the tetrahedral cation site in zinc-blende crystal, the degeneracy is partially lifted (see Fig.~\ref{FigS1}): the $^{5}$D term splits into 15-fold degenerate orbital triplet $^{5}$T$_{2}$ and 10-fold degenerate orbital doublet $^{5}$E. The Jahn-Teller distortion reduces the symmetry to D$_{2d}$ and leads to a splitting of the $^{5}$T$_{2}$ ground state into a 5-fold degenerate $^{5}$B$_{2}$ orbital singlet and a $^{5}$E orbital doublet .

The ground state orbital singlet $^{5}$B$_{2}$ is further split by the spin-orbit interaction. In a strain free crystal, it was found that the ground state splitting can be described by the spin effective Hamiltonian \cite{Vallin1974}:

\begin{eqnarray}
\label{exchange} {\cal H}_{Cr,CF}=D_0S_z^2+\frac{1}{180}F[35S_z^2-30S(S+1)S_z^2+25S_z^2]+\frac{1}{6}a[S_1^4+S_2^4+S_3^4]
\end{eqnarray}

\noindent with the Cr spin S=2 and $|D_0|\gg|a|$, $|F|$. In the model presented here, we use $a=0$ and $F=0$. The x, y, z principal axes were found to coincide with the cubic axes (1,2,3) giving rise to three identical sites, each given by (\ref{exchange}) but with the z axis of each along a different cubic axis (1,2,3). A value of D$_0\approx+30 \mu eV$ was estimated from Electron Paramagnetic Resonance (EPR) measurements in highly diluted bulk (Cd,Cr)Te \cite{Vallin1974}.

Static biaxial compressive strain in the (001) plane, as observed in self-assembled quantum dots, reduces the symmetry to D$_{2d}$ and destabilize the Cr $3d$ orbitals $d_{xz}$ and $d_{yz}$ having an electron density pointing along the $[$001$]$ axis ($z$ axis). The Cr ground state is then a 5-fold degenerated orbital singlet formed from the $d_{xy}$ orbital. It corresponds to the Jahn-Teller ground state with a tetragonal distortion along the $[$001$]$ axis \cite{Brousseau1988}.

An applied stress will also  influence the Cr spin fine structure splitting through the modification of the crystal field and the spin-orbit interaction \cite{Vallin1974}. For an arbitrary strain tensor, the general form of the Cr ground state spin effective Hamiltonian is

\begin{eqnarray}
{\cal H}_{Cr,\varepsilon}=c_1e_AS_{\theta}+c_2e_{\theta}S_{\theta}+c_3e_{\epsilon}S_{\epsilon}+c_4e_{\zeta}S_{\zeta}
+c_5(e_{\xi}S_{\xi}+e_{\eta}S_{\eta})
\end{eqnarray}

\noindent with S$_i$ defined as:

\begin{eqnarray}
S_{\theta}=S_{z}^2-\frac{1}{2}[S_{x}^2+S_{y}^2]\nonumber\\
S_{\epsilon}=\frac{1}{2}\sqrt{3}[S_{x}^2-S_{y}^2]\nonumber\\
S_{\xi}=S_{y}S_{z}+S_{z}S_{y}\nonumber\\
S_{\eta}=S_{x}S_{z}+S_{z}S_{x}\nonumber\\
S_{\zeta}=S_{x}S_{y}+S_{y}S_{x}
\end{eqnarray}

\noindent and e$_i$ defined similarly as:

\begin{eqnarray}
e_{\theta}=\varepsilon_{zz}-\frac{1}{2}[\varepsilon_{xx}+\varepsilon_{yy}]\nonumber\\
e_{\epsilon}=\frac{1}{2}\sqrt{3}[\varepsilon_{xx}-\varepsilon_{yy}]\nonumber\\
e_{\xi}=\varepsilon_{yz}+\varepsilon_{zy}\nonumber\\
e_{\eta}=\varepsilon_{xz}+\varepsilon_{zx}\nonumber\\
e_{\zeta}=\varepsilon_{xy}+\varepsilon_{yx}\nonumber\\
e_A=\varepsilon_{xx}+\varepsilon_{yy}+\varepsilon_{zz}
\end{eqnarray}

For flat self-assembled quantum dots with dominant large biaxial strain we have a strain tensor:

\begin{equation}\label{H-exc}
\mathcal{\varepsilon}_{ij} = \left(
\begin{array}{ccc}
\varepsilon_{\parallel}             &0                              &0                \\
0                                &\varepsilon_{\parallel}           &0                \\
0                                &0                              &\varepsilon_{zz}    \\
\end{array}\right)
\end{equation}

\noindent with

\begin{eqnarray}
\varepsilon_{zz}=-2\frac{C_{11}}{C_{12}}\varepsilon_{\parallel}
\end{eqnarray}

\noindent where C$_{11}\approx$ 5.4 10$^{10}Pa$ and C$_{12}\approx$ 3.7 10$^{10}Pa$ are the elastic constants of CdTe \cite{Greenough1973}. For this strain configuration, the Cr fine structure is controlled by the spin-lattice coupling coefficients c$_1$ (symmetric coefficient) and c$_2$ (tetragonal coefficients). Strain-coupling coefficients estimated from EPR measurements in bulk Cr doped CdTe are listed in table \ref{table1}.

\begin{table}[htb] \centering
\caption{Values for spin to strain coupling coefficients of Cr in bulk CdTe (in $meV$) extracted from ref.\cite{Vallin1974}.}
\renewcommand{\arraystretch}{1.0}
\begin{tabular}{p{1.5cm}p{1.5cm}p{1.5cm}p{1.5cm}p{1.5cm}}
\hline\hline
c$_{1}$ & c$_{2}$ & c$_{3}$  & c$_{4}$  & c$_{5}$ \\
-0.25$\pm$2 & +4.9 $\pm$2& -1.25$\pm$0.5 & +4.9$\pm$2 & +3.7$\pm$1.25 \\
\hline\hline
\end{tabular}
\label{table1}
\end{table}

For a pure CdTe layer lattice matched on ZnTe, we have $\varepsilon_{\parallel}=(a_{ZnTe}-a_{CdTe})/a_{CdTe}\approx-5.8\%$ with $a_{CdTe}=6.48{\AA}$ and $a_{ZnTe}=6.10{\AA}$. The strain controlled part of the spin Hamiltonian ${\cal H}_{Cr,\varepsilon}$ becomes:

\begin{eqnarray}
{\cal H}_{Cr,\varepsilon_\parallel}= \frac{3}{2}\varepsilon_{\parallel}[2c_1(1-\frac{C_{12}}{C_{11}})-c_2(1+2\frac{C_{12}}{C_{11}})]S_z^2=D_0S_z^2
\end{eqnarray}

\noindent where we can estimate D$_0\approx$ 1$\pm$0.6 meV from the values of the spin-strain coupling coefficients in CdTe (table \ref{table1}). However one should note the quantum dots could be partially relaxed and may contain a significant amount of Zn. We were not able to find spin to strain coupling coefficients for Cr in ZnTe and (Cd,Zn)Te alloy in literature. A value of D$_0\approx$+280$\mu$eV, much larger than for CdTe, was however estimated in strain-free Cr-doped bulk ZnTe \cite{Vallin1974}. Larger spin-strain coupling coefficients could then expected for Cr in ZnTe and (Cd,Zn)Te alloys.

Finally, an anisotropy of the strain in the quantum dot plane (001) with principal axis along $[$010$]$ or $[$100$]$ axes ($\varepsilon_{xx}\neq\varepsilon_{yy}$ and $\varepsilon_{xy}$=$\varepsilon_{yx}$=0) would affect the Cr fine structure through the tetragonal coefficient c$_3$. This coupling can be described by an additional term in the spin-strain Hamiltonian

\begin{eqnarray}
{\cal H}_{Cr,\varepsilon\perp}= \frac{3}{4}c_3(\varepsilon_{xx}-\varepsilon_{yy})(S_x^2-S_y^2)=E(S_x^2-S_y^2)
\end{eqnarray}

This anisotropy term E couples Cr spin states separated by two units and in particular S$_z$=+1 to S$_z$=-1 which are initially degenerated. It could be exploited to induce a large strain mediated coherent coupling between a mechanical oscillator and the Cr spin \cite{Ovar2014}.

\section{Optical transitions in a Cr-doped quantum dot}

The complete electron-hole-Cr Hamiltonian in self-assembled quantum dot (${\cal H}_{X-Cr}$) can be separated into six parts:

\begin{eqnarray}
\label{X-Cr} {\cal H}_{X-Cr}={\cal H}_{Cr,\varepsilon}+{\cal H}_{c-Cr}+{\cal H}_{mag}+{\cal H}_{e-h}+{\cal H}_{band}+{\cal H}_{scat}
\end{eqnarray}

${\cal H}_{Cr,\varepsilon}$ describes the fine structure of the Cr atom and its dependence on local strain as presented in the previous section.

${\cal H}_{c-Cr}$ describes the coupling of the electron and hole with the Cr spin. It reads

\begin{eqnarray}
\label{c-Cr} {\cal H}_{c-Cr}= I_{eCr}\overrightarrow{S}.\overrightarrow{\sigma}+I_{hCr}\overrightarrow{S}.\overrightarrow{J}
\end{eqnarray}

\noindent with $I_{eCr}$ and $I_{hCr}$ the exchange integrals of the electron ($\overrightarrow{\sigma}$) and hole ($\overrightarrow{J}$) spins with the Cr spin ($\overrightarrow{S}$).

An external magnetic field couples via the standard Zeeman terms to both the Cr spin and carriers spins and a diamagnetic shift of the electron-hole pair can also be included resulting in

\begin{eqnarray}
\label{cmag3} {\cal H}_{mag}=g_{Cr}\mu_B\overrightarrow{B}.\overrightarrow{S}+g_{e}\mu_B\overrightarrow{B}.\overrightarrow{\sigma}+g_{h}\mu_B\overrightarrow{B}.\overrightarrow{J}+\gamma B^2
\end{eqnarray}

The electron-hole exchange interaction, ${\cal H}_{e-h}$, contains the short range and the long range parts. The short range contribution is a  contact interaction which induces a splitting $\delta_0^{sr}$ of the bright and dark excitons and, in the reduced symmetry of a zinc-blend crystal ($T_d$), a coupling  $\delta_2^{sr}$ of the two dark excitons. The long range part also contributes to the bright-dark splitting by an energy $\delta_0^{lr}$. In quantum dots with C$_{2v}$ symmetry (ellipsoidal flat lenses for instance \cite{Zielinski2015}) the long range part also induce a coupling $\delta_1$ between the bright excitons. Realistic self-assembled quantum dots have symmetries which can deviate quite substantially from the idealized shapes of circular or ellipsoidal lenses. For a $C_{s}$ symmetry (truncated ellipsoidal lens), additional terms coupling the dark and the bright excitons have to be included in the electron-hole exchange Hamiltonian. Following Ref. \cite{Zielinski2015}, the general form of the electron-hole exchange Hamiltonian in the heavy-hole exciton basis $|+1\rangle$, $|-1\rangle$, $|+2\rangle$, $|-2\rangle$ for a low symmetry quantum dot (C$_s$) is

\begin{equation}\label{Heh}
\mathcal{H}_{e-h} =\frac{1}{2} \left(
\begin{array}{cccc}
-\delta_0                               &e^{i\pi/2}\delta_1              &e^{i\pi/4}\delta_{11}        &-e^{i\pi/4}\delta_{12}\\
e^{i\pi/2}\delta_1                      &-\delta_0                       &e^{-i\pi/4}\delta_{12}       &-e^{-i\pi/4}\delta_{11}\\
e^{-i\pi/4}\delta_{11}                  &e^{i\pi/4}\delta_{12}           &\delta_0                     &\delta_2\\
-e^{-i\pi/4}\delta_{12}                 &-e^{i\pi/4}\delta_{11}          &\delta_2                     &\delta_0\\
\end{array}\right)
\end{equation}

The terms $\delta_{11}$ and $\delta_{12}$, not present in symmetry C$_{2v}$, give an in-plane dipole moment to the dark excitons \cite{Bayer2002}. The term $\delta_{12}$ which couples $|\pm1\rangle$ and $|\mp2\rangle$ excitons respectively is responsible for the dark-bright anti-crossing observed on the low energy side of the emission of Cr-doped quantum dots.

\begin{table*}[t] \centering
\caption{Values of the parameters used in the model of Cr-doped CdTe/ZnTe quantum dot presented in figure 3 of the main text. The value of the parameters not listed in the table is 0. The chosen values are typical for CdTe/ZnTe quantum dots and can be compared with parameters extracted from Mn-doped quantum dots \cite{Besombes2014,Leger2007}.}
\renewcommand{\arraystretch}{1.0}
\begin{tabular}{p{0.9cm}p{0.9cm}p{0.9cm}p{0.9cm}p{0.9cm}p{0.9cm}p{0.9cm}p{1.0cm}p{0.9cm}p{0.9cm}p{0.9cm}p{0.9cm}p{1.3cm}p{0.9cm}p{0.9cm}}
\hline\hline
I$_{eCr}$ & I$_{hCr}$ & $\delta_0$ & $\delta_1$ & $\delta_{12}$ & $\delta_{11}$ & $\frac{\delta_{xz}}{\Delta_{lh}}$ & $\frac{\delta_{xx,yy}}{\Delta_{lh}}$ & $D_0$ & $g_{Cr}$ & $g_{e}$ & $g_{h}$ & $\gamma$ & $\eta$ & $T_{eff}$ \\
$\mu eV$ & $\mu eV$ & $meV$ & $\mu eV$ & $\mu eV$ & $\mu eV$ &  & & $meV$&  & &  & $\mu eV/T^2$ & $\mu eV$ & K \\
\hline
-70 & -280 & -1 & 250 & 150 & 50 & 0.05 & 0.05& 2.5 & 2 & -0.7 & 0.4 & 1.5 & 25 & 25 \\
\hline\hline
\end{tabular}
\label{table2}
\end{table*}

The band Hamiltonian, ${\cal H}_{band}=E_g+\mathcal{H}_{vbm}$, stands for the energy of the electrons (i.e. the band gap energy E$_g$), and the heavy-holes (hh) and light-holes (lh) energies ($\mathcal{H}_{vbm}$) \cite{Leger2007,Besombes2014}. To describe the influence of a reduced symmetry of the quantum dot on the valence band, we considered here the four lowest energy hole states $|J,J_z\rangle$ with
angular momentum $J=3/2$. A general form of Hamiltonian describing the influence of shape or strain anisotropy on the valence band structure can be written in the basis ($|\frac{3}{2},+\frac{3}{2}\rangle,|\frac{3}{2},+\frac{1}{2}\rangle,|\frac{3}{2},-\frac{1}{2}\rangle,|\frac{3}{2},-\frac{3}{2}\rangle$) as:

\begin{equation}\label{Hvbm}
\mathcal{H}_{vbm} = \left(
\begin{array}{cccc}
0                                &-Q                                           &R                                   &0\\
-Q^*                             &\Delta_{lh}                                  &0                                   &R\\
R^*                              &0                                            &\Delta_{lh}                         &Q\\
0                                &R^*                                          &Q^*                                 &0\\
\end{array}\right)
\end{equation}

\noindent with

\begin{eqnarray}
\label{exchange3} Q=\delta_{xz}-i\delta_{yz}; R=\delta_{xx,yy}-i\delta_{xy}
\end{eqnarray}

Here, R describes the heavy-hole / light-hole mixing induced by an anisotropy in the quantum dot plane $xy$ and Q takes into account an asymmetry in the plane containing the quantum dot growth axis $z$. The reduction of symmetry can come from the shape of the quantum dot (Luttinger Hamiltonian) or the strain distribution (Bir and Pikus Hamiltonian). $\Delta_{lh}$ is the splitting between lh and hh which is controlled in quantum dots both by the in-plane biaxial strain and the confinement.

Considering only an in-plane anisotropy (Q=0), it follows from (\ref{Hvbm}) that the valence band mixing couples the heavy-holes $J_z=\pm3/2$ and the light-holes $J_z=\mp1/2$ respectively. For such mixing, the isotropic part of the short range exchange interaction, which can be written in the form $2/3\delta_0^{sr}(\overrightarrow{\sigma}.\overrightarrow{J})$, couples the two bright excitons. This mixing is also responsible for a weak $z$-polarized dipole matrix element of the dark excitons coming from the light-hole part of the hole wave function.

A deformation in a vertical plane (Q term) couples the heavy-holes $J_z=\pm3/2$ and the light-holes $J_z=\pm1/2$ respectively. In this case, the short range electron-hole exchange interaction couples $|+1\rangle$ and $|+2\rangle$ exciton on one side and $|-1\rangle$ and $|-2\rangle$ exciton on the other side.

For a general description and as it was observed in Mn-doped quantum dots \cite{Besombes2005,Trojnar2013,Besombes2014}, we can also take into account the perturbation of the wave function of the exciton in the initial state of the optical transition by the hole-Cr exchange interaction. This perturbation depends on the value of the exchange energy between the Cr spin S$_z$  and the hole spin J$_z$ and can be represented, using second order perturbation theory, by an effective spin Hamiltonian \cite{Besombes2005,Trojnar2013,Besombes2014}

\begin{eqnarray}
{\cal H}_{scat}=-\eta S_z^2
\end{eqnarray}

\noindent with $\eta>0$.

Using the Hamiltonian of the excited state ${\cal H}_{X-Cr}$ and the Hamiltonian of the ground state

\begin{eqnarray}
{\cal H}_{Cr}={\cal H}_{Cr,\varepsilon}+g_{Cr}\mu_B\overrightarrow{B}.\overrightarrow{S}
\end{eqnarray}

\noindent we can compute the spectrum of a quantum dot containing a Cr atom. The occupation of the X-Cr levels is described by an effective spin temperature T$_{eff}$ and the optical transitions probabilities are obtained calculating the matrix elements $|\langle S_z|X,S_z\rangle|^2$ where $X$ and S$_z$ stands for the 8 possible exciton states and the Cr spin respectively. The resulting photoluminescence spectra calculated with the parameters listed in table \ref{table2} are presented in figure 3 of the main text.

\section{Complementary data on photoluminescence of Cr-doped quantum dots}

\begin{figure}
   \begin{minipage}[c]{.48\linewidth}
      \centering{\includegraphics[width=2.75in]{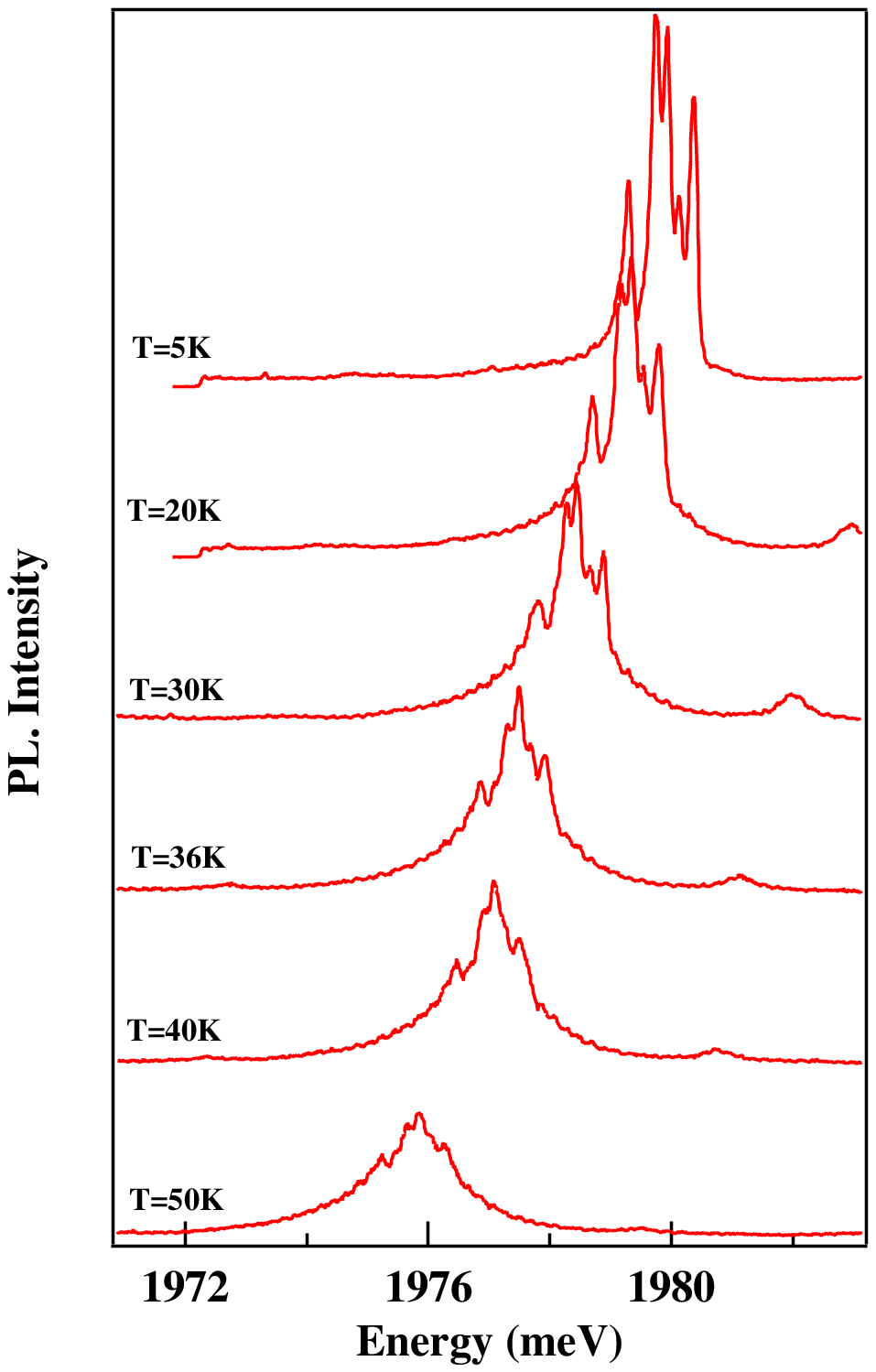}
      \caption{Temperature dependence of the photoluminescence of X-Cr in a Cr-doped QD (QD4 in the main text).}
      \label{FigS2}}
   \end{minipage} \hfill
   \begin{minipage}[c]{.48\linewidth}
      \centering{\includegraphics[width=2.5in]{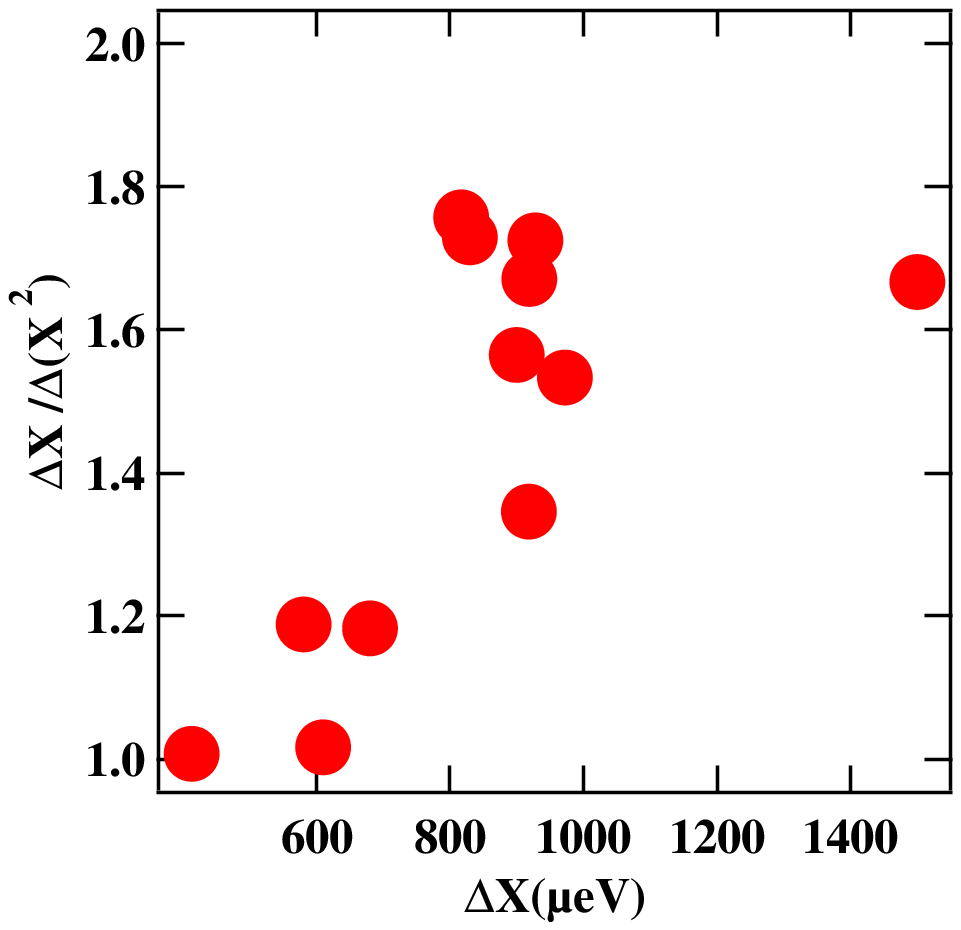}
      \caption{Ratio of the overall splitting (i.e energy difference between the two extreme bright exciton lines) of X$^2$-Cr and X-Cr as a function of the splitting of X-Cr for 12 different Cr-doped QDs.}
      \label{FigS3}}
   \end{minipage}
\end{figure}

The temperature dependence of the X-Cr emission in QD4 at zero magnetic field is presented in figure \ref{FigS2}. With the increase of the temperature, we observe a significant line broadening induced by the interaction with acoustic phonons. In the investigated temperature range where we still obtain a significant PL intensity and resolved PL lines (below T=50K), no contribution of the S$_z$=$\pm$2 Cr spins states are observed in the emission of the exciton.

Figure \ref{FigS3} presents some statistics on the values of the overall energy splitting of X-Cr ($\Delta X$) and X$^2$-Cr ($\Delta X^2$) obtain on 12 Cr-doped quantum dots. For most of the investigated dots, $\Delta X$ remains below 1 meV and $\Delta X^2$ is also smaller than $\Delta X$ suggesting an interaction between the Cr spin and the biexciton.

\end{document}